# Human Indignity:
## From Legal AI Personhood to Selfish Memes

**Roman V. Yampolskiy**
Computer Engineering and Computer Science
University of Louisville
roman.yampolskiy@louisville.edu

**Abstract**
It is possible to rely on current corporate law to grant legal personhood to Artificially Intelligent (AI) agents. In this paper, after introducing pathways to AI personhood, we analyze consequences of such AI empowerment on human dignity, human safety and AI rights. We emphasize possibility of creating selfish memes and legal system hacking in the context of artificial entities. Finally, we consider some potential solutions for addressing described problems.

> *"The Machine is not an it to be animated, worshiped, and dominated. The machine is us, our process, an aspect of our embodiment. We can be responsible for machines; they do not dominate or threaten us. We are responsible for boundaries; we are they."* - Donna Haraway

## 1. Introduction to AI Personhood

Debates about rights are frequently framed around the concept of legal personhood, which is granted not just to human beings but also to some non-human entities, such as firms, corporations or governments. Legal entities, aka legal persons are granted certain privileges and responsibilities by the jurisdictions in which they are recognized, and many such rights are not available to non-person agents. Attempting to secure legal personhood is often seen as a potential pathway to get certain rights and protections for animals [1], fetuses [2], trees, rivers [3] and artificially intelligent (AI) agents [4]. It is commonly believed that a court ruling or a legislative action is necessary to grant personhood to a new type of entity, but recent legal literature [5-8] suggests that loopholes in the current law may permit granting of legal personhood to currently existing AI/software without having to change the law or persuade any court.

LoPucki [6] in his paper on Algorithmic Entities cites Bayern [7, 8] and his work on conferring legal personhood on AI by putting it in charge of LLC[1]. "Professor Shawn Bayern demonstrated

---
[1] "Bayern specifies this chain of events as capable of establishing the link: (1) [A]n individual member creates a member-managed LLC, filing the appropriate paperwork with the state; (2) the individual (along, possibly, with the LLC, which is controlled by the sole member) enters into an operating agreement governing the conduct of the LLC; (3) the operating agreement specifies that the LLC will take actions as determined by an autonomous system,

that anyone can confer legal personhood on an autonomous computer algorithm merely by putting it in control of a limited liability company (LLC). The algorithm can exercise the rights of the entity, making them effectively rights of the algorithm. The rights of such an algorithmic entity (AE) would include the rights to privacy, to own property, to enter into contracts, to be represented by counsel, to be free from unreasonable search and seizure, to equal protection of the laws, to speak freely, and perhaps even to spend money on political campaigns. Once an algorithm had such rights, Bayern observed, it would also have the power to confer equivalent rights on other algorithms by forming additional entities and putting those algorithms in control of them."[2] [6].

Other legal pathways to obtain legal personhood have been suggested and analyzed in the literature [4-12], but details of such legal "hacking" are beyond the scope of this paper. We are simply interested in understanding the impact of granting personhood to AI on human dignity [13] and safety. With appearance of decentralized autonomous organizations [14], such as the DAO [15], these questions are as pressing as ever.

## 2. Selfish Memes

In his book, The Selfish Gene, Dawkins [16] talks about genes as the driving payload behind evolution, with animal bodies as vehicles for the gene to accomplish its goals in the world. He also introduced a new concept of a *meme*, or viral idea competing for dominance in human minds, inspired by the similarities between natural and cultural evolution. Advent of algorithmic entities would make it possible to explicitly add a memetic payload to a legal entity, resulting in what we will call the Selfish Meme. Corporations are selfish entities with the goal of maximizing shareholder profit, with AI in charge of such an entity any idea can be codified in an algorithm and added as the driving force behind the corporation's decision making. At a higher level of abstraction this could produce selfish cryptocurrencies.

We already see something similar from B-corps or for Benefit Corporations, which attempt to create some social good in addition to profit. However, such memetic payload doesn't have to be strictly beneficial, in practice it could be any ideology, set of beliefs or values. For example, it would be possible to codify tenants of a particular religion (ex. Islam), economic philosophy (communism), moral theory (Utilitarianism) or something silly but potentially dangerous like a Paperclip Maximizer [17] or Pepe meme [18], encode them in an algorithm and put that algorithm in charge of a corporation, which could eventually take over the world and inforce its memetic payload on the world. Via orthogonality thesis [19] we can see that few if any limitations exists on the potential memetic payload, it could be a marketing campaign, an uploaded animal or human mind, our constitution and the complete set of laws or a computer virus. Evolutionary competition would appear between such entities leading to adversarial practices [20] and perhaps hostile takeovers not just in a legal but also in a computer science sense, with hacking and replacement of corporation's selfish meme with another payload, being a real possibility. We live in the world where computer viruses may have rights. It would also not be surprising if establishment of

---

specifying terms or conditions as appropriate to achieve the autonomous system's legal goals; (4) the sole member withdraws from the LLC, leaving the LLC without any members. The result is potentially a perpetual LLC—a new legal person—that requires no ongoing intervention from any preexisting legal person in order to maintain its status. AEs would not be confined to cyberspace. An AE could act offline by contracting online with humans or robots for offline services. Bayern uses an algorithm that operates a Bitcoin vending machine business to illustrate."
[2] See original article for footnotes, which have been removed to improve readability of quotes.

corporations with custom memetic payload become available as a service. Such corporations may also attempt to gain membership in different organizations and partnerships for example the recently formed Partnership on AI[3], in order to influence it from within.

## 3. Human Indignity

The process for getting legal rights for AI, described above, doesn't specify any minimal intelligence/capability for the AI involved, meaning it could be just a few "if" statements, a random decision generator or an emulation of an ameba[4]. To grant most if not all human rights to a cockroach, for example, would be an ultimate assault on human dignity (but it may make some people happy [21]). This could be potentially done as an art project or in protest of unequal treatment of all humans by some human rights activists. We already witnessed an example of such indignity and subsequent outrage from many feminist scholars [22] on the news of Sophia the robot getting citizenship in the Saudi Arabia, a country notorious for unequal treatment of women. Will self-driving cars be allowed on the roads before women are?[5] As a result of legal personhood and granting of associated rights, some humans will have less rights than trivial (non-intelligent) software and robots, a great indignity and discriminatory humiliation. For example, certain jurisdictions limit rights of their citizens, such as a right to free speech, freedom or religious practice, or expression of sexuality, but AIs with legal personhood in other jurisdictions would be granted such rights.

If, on the other hand, AIs are going to become more intelligent than humans, the indignity for humanity would come from being relegated to an inferior place in the world, being outcompeted in the workplace and all other domains of human interest [23, 24]. AI led corporations would be in position to fire their human workers. This would possibly lead to deteriorating economic and living conditions, permanent unemployment and potentially reduction in rights, not to mention further worsening of the situation including to the level of existential catastrophe (extermination) [25].

The precedent of AI obtaining legal personhood via the corporate loophole may catalyze legislative granting of equal rights to artificially intelligent agents as a matter of equal treatment, leading to a number of indignities for the human population. Since software can reproduce itself almost indefinitely, they would quickly make human suffrage inconsequential, if given civil rights [26] leading to the loss of self-determination for people. Such loss of power would likely lead to the redistribution of resources from humanity to machines as well as possibility of AIs serving as leaders, presidents, judges, jurors and even executioners. We will see military AIs targeting human populations and deciding on their own targets and acceptable collateral damage. They may not necessarily subscribe to the Geneva Convention and other rules of war. Torture, genocide and nuclear war may become options to consider to reach desired goals.

As AIs' capabilities and dominance grow, they would likely self-grant special (super) rights to emphasize their superiority to people, while at the same time removing or at least reducing human rights (ex. 2$^{nd}$ amendment, 1$^{st}$ amendment, reproductive rights in the sense of the right to reproduce at all, aka, 0-child policy, Convention on human rights, etc.) while justifying doing so by our relative "feeblemindedness". A number of scholars [27-29] today work on developing reasons for justifying granting of rights to AIs, perhaps one day those reasons will be useful while we are begging to keep some of ours.

---

[3] https://www.partnershiponai.org/
[4] Same legal loophole could be used to grant personhood to animals or others with inferior rights.
[5] As of June 24, 2018 and after this was written, women were permitted to drive in Saudi Arabia.

## 4. Legal-System Hacking

Corporations can act as their own lawyers while representing themselves in the court of law, including performing all functions of a human lawyer, such as sue and be sued. Artificial superintelligence in charge of a corporation can act as a super-lawyer capable of finding novel loopholes in our laws (zero-day law exploits), engaging in frivolous litigation (DOS-style litigation attacks), patent filing and trolling, and smart-contract fallibility detection [30]. Our laws are complex, ambiguous and too numerous to be read by any single person, with USA tax-code alone approaching 4,000 (or 75,000 if you include IRS explanations, regulations and rulings) pages, making it perfect for AI to exploit by both finding flaws in existing contracts and drafting contracts with hard-to-detect backdoors. A meeting of the minds between a human and superintelligence is unlikely to be achievable.

It is also likely that computational legal language [31] and smart contracts [32] will come to replace our current legal code making it inaccessible to human lawyers due to it computational complexity, size and unnatural jargon further contributing to our second-class citizen status and indignity. This would happen simultaneously with the current trend of digitizing judiciary system and civil engagement as illustrated by Korean e-judiciary [33] and Estonian e-residency program [34], trends which while providing short-term convenience to people, give long-term advantage to the machines. This seems to be a part of a larger trend of society moving to Algocracy – rule by algorithm, there code is law [35]. Furthermore, due to its comparative advantage in large scale surveillance and data mining AI would be able to uncover human illegal behavior, and as most have broken some law (ex. tax evasion, speeding, obscure laws, etc.), bring legal action or threat of legal actions against everyone. Similar blackmail and reporting could happen in the business environment with AI also enjoying existing whistleblower protections.

## 5. Human Safety

A lot has been published on different risks associated with advancement of artificial intelligence [36-43], but less specifically on dangers from AI controlled corporate entities. Nothing in our current laws would prevent formation of a malevolent corporation (or corporate virus) with memetic payload of subjugating or exterminating humanity through legal means and political influence. In addition to legal enslavement of people via below living-wage salary, such corporations could support legal change in minimum wage and pension laws as well as provide opposition to wealth redistribution and Universal Basic Income/Universal Basic Assets [44, 45]. This is particularly easy to accomplish because of Supreme Court decision in Citizens United VS FEC [46], permitting unrestricted donations from corporations to politicians under the guise of free speech, making it possible to convert financial wealth to political power.

This leads us to recognize an additional existential risk (X-risk) [47], from extreme wealth. Wealth inequality is already recognized as a problem for democratic institutions [48], but super-rich corporate AIs (dollar trillionaires) would take that problem to the next level. They could accumulate such unprecedented levels of wealth via unfair business practices such as predatory pricing, having access to free physical and cognitive labor from direct control of automation, permitting them to undermine completion and achieve monopoly status in multiple domains and other multiple resources. Additionally, such entities could engage in super-long term investment getting compound interest for hundreds of years. For example, a million dollars invested for 150 years at the same rate of return as observed over the last hundred years would grow to 1.6 trillion inflation-unadjusted dollars, creating super-rich artificial entities.

If EAs become intellectually indistinguishable from people, meaning could pass an unrestricted Turing Test [49] their capacity to self-replicate could be used to drain resources from legitimate corporations, for example via click-fraud [50, 51] from Google. Also, they will be able to create their own super successful companies with alternative populations comprised of billions of EA users indistinguishable from real people but paid for by advertisers as if they were genuine clients.

Super-rich would be able to work within and outside the law, using donations or bribes to influence politicians, as well as directly breaking the law and simply paying fines for such actions. Because corporations can create other corporations it would become possible to establish a legally independent suicidal corporation, which is willing to accomplish any legal or illegal goal of a originator corporation and after that cease to exist, permitting avoidance of any responsibility by the original algorithm entity. With appearance of dark-web assassin markets financed through anonymous crypto-payments [52] the power of the super-rich can't be effectively fought against without endangering personal safety and security. At the least, the super-rich have the power to ruin someone's life financially, socially and professionally if direct termination is not preferred. Politicians financially backed by algorithmic entities would be able to take on legislative bodies, impeach president and help to get figureheads appointed to the Supreme Court. Such human figureheads could be used to obtain special (super-rights) for AIs or at least expansion of corporate rights. It also may become possible to exercise direct control over human figureheads via advanced Computer-Brain Interfaces (CBI) permitting AIs unrestricted manipulation of a human body, essentially turning them into meat avatars, another source of indignity.

LoPucki provides a detailed list of reasons a human may set up an AE [6]: "

1. *Terrorism.* An initiator could program an AE to raise money to finance terrorism or to directly engage in terrorist acts. It could be programmed for genocide or general mayhem.
2. *Benefits.* An initiator could program an AE to provide direct benefits to individuals, groups, or causes. …
3. *Impact.* An initiator could program an AE to achieve some specified impact on the world. The goals might range all of the way from traditional philanthropy to pure maliciousness. …
4. *Curiosity.* An initiator might launch an AE simply out of curiosity. Initiators have sometimes devoted substantial time and money to launch computer viruses from which they could derive no monetary benefit. …
5. *Liability avoidance.* Initiators can limit their civil and criminal liability for acts of their algorithms by transferring the algorithms to entities and surrendering control at the time of the launch. For example, the initiator might specify a general goal, such as maximizing financial return, and leave it to the algorithm to decide how to do that." [6].

What makes artificial entities particularly difficult to control, compete against and overall dangerous is that they enjoy a number of super-properties natural persons do not have. They are effectively immortal, non-physical, optimizable, and get more capable with time as they accumulate computational and financial resources. They are much more flexible in terms of their energy, temperature, storage needs as compared to biological entities. From the legal point of view, they can't be legally punished, or terminated, and are generally not subject to law enforcement as our judicial system is not set up for such entities [53]. Neither prisons, nor corporal nor capital punishment is applicable to algorithmic entities.

LoPucki also analyzes a number of similar and concerning properties of AEs, which differentiate them from natural persons and give them a strategic advantage [6]: "

   a. **Ruthlessness** Unless programmed to have them, AEs will lack sympathy and empathy. Even if the AEs are fully capable of understanding the effects of their actions on humans, they may be indifferent to those effects. As a result, AEs will have a wider range of options available to them than would be available to even the most morally lax human controller. An AE could pursue its goals with utter ruthlessness. Virtually any human controller would stop somewhere short of that, making the AE more dangerous.
   b. **Lack of Deterrability**
   Outsiders can more easily deter a human-controlled entity than an AE. For example, if a human-controlled entity attempts to pursue an illegal course of action, the government can threaten to incarcerate the human controller. If the course of action is merely abhorrent, colleagues, friends, and relatives could apply social pressures. AEs lack those vulnerabilities because no human associated with them has control. As a result, AEs have greater freedom to pursue unpopular goals using unpopular methods. In deciding to attempt a coup, bomb a restaurant, or assemble an armed group to attack a shopping center, a human-controlled entity puts the lives of its human controllers at risk. The same decisions on behalf of an AE risk nothing but the resources the AE spends in planning and execution.
   c. **Replication**
   AEs can replicate themselves quickly and easily. If an AE's operations are entirely online, replication may be as easy as forming a new entity and electronically copying an algorithm. An entity can be formed in some jurisdictions in as little as an hour and for as little as seventy dollars. … Easy replication supports several possible strategies. First, replication in a destination jurisdiction followed by dissolution of the entity in the original jurisdiction may put the AE beyond the legal reach of the original jurisdiction. … Second, replication can make an AE harder to destroy. For example, if copies of an AE exist in three jurisdictions, each is a person with its own rights. A court order revoking the charter of one or seizing the assets of another would have no effect on the third." [6].

Such AEs would be far less scrupulous about running casinos, or brothels, or selling drugs all business, which while potentially legal may have significant impact on human dignity.

With development of advanced robot bodies it will become possible for AEs to embody themselves to more fully participate in the world and to directly perform physical actions which otherwise require multiple levels of indirect control. An EA can potentially be running on a humanoid robot or a self-driving car, or a flying drone or any sufficiently powerful embedded processor or cloud service. This by extension would permit memetic payloads to acquire bodies resulting in the next level of evolutionary competition, in which a computer virus meme or a biological viral gene may propagate through a human-like body. If quality of such humanoid robots is high enough to pass a Total Turing Test [54] it would become impossible to tell between a natural and artificial people likely leading to the violation of the Turing's Red Flag law [55]. Consequently, people would have an option to continue to exist and influence the world after their death via embodied representative algorithms. At the same time, autonomous corporation would have an option to replace human employees with identical but controlled clones. Similar analysis can be performed for virtual worlds and avatar bodies.

## 6. Conclusions

In this paper, we looked at a number of problems, which AI personhood can cause as well as direct impact on human dignity from such legal recognition. The question before us: is there anything we can do to avoid such dehumanizing future? While some solutions may be possible in theory, it does not mean that they are possible in practice. Changing the law to explicitly exclude AIs from becoming legal entities may be desirable but unlikely to happen in practice, as that would require changing existing corporate law across multiple jurisdictions and such major reforms are unlikely to pass. Perhaps it would be helpful to at least standardize corporate law across multiple jurisdictions, but that is likewise unlikely to happen. Similarly, laws regarding maximum wealth levels, to prevent accumulation of extreme wealth have no chance of passing and would be easily bypassed by clever AIs if introduced.

Overall, it is important to realize that just like hackers attack computer systems and discover bugs in the code, machines will attack our legal systems and discover bugs in our legal code and contracts. For every type of cybersecurity attack, a similar type of attack will be discovered in the legal domain. Number of such attacks and their severity will increase proportionate to the capabilities of AIs. To counteract such developments, we need to establish, understand and practice Legal Safety the same way we do cybersecurity. The only good news is that consequences from successful legal attacks are likely to be less severe compared to direct threats we will face from malevolent superintelligences.

## Acknowledgments
The author is grateful to Elon Musk and the Future of Life Institute and to Jaan Tallinn and Effective Altruism Ventures for partially funding his work on AI Safety.

## References

1. Varner, G.E., *Personhood, ethics, and animal cognition: Situating animals in Hare's two level utilitarianism*. 2012: Oxford University Press.
2. Schroedel, J.R., P. Fiber, and B.D. Snyder, *Women's Rights and Fetal Personhood in Criminal Law*. Duke J. Gender L. & Pol'y, 2000. **7**: p. 89.
3. Gordon, G.J., *Environmental Personhood*. Colum. J. Envtl. L., 2018. **43**: p. 49.
4. Chopra, S. and L. White. *Artificial agents-personhood in law and philosophy*. in *Proceedings of the 16th European Conference on Artificial Intelligence*. 2004. IOS Press.
5. Solum, L.B., *Legal personhood for artificial intelligences*. NCL Rev., 1991. **70**: p. 1231.
6. LoPucki, L.M., *Algorithmic Entities*. Washington University Law Review, 2018. **95(4)**.
7. Bayern, S., *The Implications of Modern Business–Entity Law for the Regulation of Autonomous Systems*. European Journal of Risk Regulation, 2016. **7**(2): p. 297-309.
8. Bayern, S., *Of Bitcoins, Independently Wealthy Software, and the Zero-Member LLC*. Northwestern University Law Review, 2013. **108**: p. 1485.
9. Andrade, F., et al., *Contracting agents: legal personality and representation*. Artificial Intelligence and Law, 2007. **15**(4): p. 357-373.
10. Calverley, D.J., *Imagining a non-biological machine as a legal person*. Ai & Society, 2008. **22**(4): p. 523-537.
11. Teubner, G., *Rights of non-humans? Electronic agents and animals as new actors in politics and law*. Journal of Law and Society, 2006. **33**(4): p. 497-521.
12. Dan-Cohen, M., *Rights, persons, and organizations: A legal theory for bureaucratic society*. Vol. 26. 2016: Quid Pro Books.

13. Bostrom, N., *In defense of posthuman dignity.* Bioethics, 2005. **19**(3): p. 202-214.
14. Dilger, W. *Decentralized autonomous organization of the intelligent home according to the principle of the immune system.* in *Systems, Man, and Cybernetics, 1997. Computational Cybernetics and Simulation., 1997 IEEE International Conference on*. 1997. IEEE.
15. DuPont, Q., *Experiments in algorithmic governance: A history and ethnography of "The DAO," a failed decentralized autonomous organization*, in *Bitcoin and Beyond*. 2017, Routledge. p. 157-177.
16. Dawkins, R., *The Selfish Gene*. 1976, New York City: Oxford University Press.
17. Yudkowsky, E., *Intelligence explosion microeconomics.* Machine Intelligence Research Institute, accessed online October, 2013. **23**: p. 2015.
18. Mele, C., *Pepe the Frog Meme Listed as a Hate Symbol.* The New York Times. Retrieved from http://www. nytimes. com/2016/09/28/us/pepe-the-frog-is-listed-as-a-hate-symbol-by-the-anti-defamation-league. html, 2016.
19. Bostrom, N., *The superintelligent will: Motivation and instrumental rationality in advanced artificial agents.* Minds and Machines, 2012. **22**(2): p. 71-85.
20. Ramamoorthy, A. and R. Yampolskiy, *Beyond MAD?: the race for artificial general intelligence.* ICT Discoveries, Special Issue No, 2018. **1**.
21. Tomasik, B., *The importance of insect suffering. Essays on Reducing Suffering*, in *Reducing Suffering*. 2016: Available at: reducing-suffering.org/the-importance-of-insect-suffering.
22. Kanso, H., *Saudi women riled by robot with no hjiab and more rights than them.* Nevember 1, 2017: https://www.reuters.com/article/us-saudi-robot-citizenship/saudi-women-riled-by-robot-with-no-hjiab-and-more-rights-than-them-idUSKBN1D14Z7.
23. Bostrom, N., *Superintelligence: Paths, dangers, strategies*. 2014: Oxford University Press.
24. Yampolskiy, R.V., *Artificial Superintelligence: a Futuristic Approach*. 2015: Chapman and Hall/CRC.
25. Pistono, F. and R.V. Yampolskiy. *Unethical Research: How to Create a Malevolent Artificial Intelligence*. in *25th International Joint Conference on Artificial Intelligence (IJCAI-16). Ethics for Artificial Intelligence Workshop (AI-Ethics-2016)*. 2016.
26. Yampolskiy, R.V., *Artificial intelligence safety engineering: Why machine ethics is a wrong approach*, in *Philosophy and Theory of Artificial Intelligence*. 2013, Springer. p. 389-396.
27. Guo, S. and G. Zhang, *Robot Rights.* Science, February 13, 2009. **323**: p. 876.
28. Coeckelbergh, M., *Robot rights? Towards a social-relational justification of moral consideration.* Ethics and Information Technology, 2010. **12**(3): p. 209-221.
29. Gunkel, D., *The Other Question: The Issue of Robot Rights.* Sociable Robots and the Future of Social Relations: Proceedings of Robo-Philosophy 2014, 2014. **273**: p. 13.
30. Yampolskiy, R.V., *What are the ultimate limits to computational techniques: verifier theory and unverifiability.* Physica Scripta, 2017. **92**(9): p. 093001.
31. Wolfram, S., *Computational Law, Symbolic Discourse and the AI Constitution*. 2016: http://blog.stephenwolfram.com/2016/10/computational-law-symbolic-discourse-and-the-ai-constitution/
32. Christidis, K. and M. Devetsikiotis, *Blockchains and smart contracts for the internet of things.* Ieee Access, 2016. **4**: p. 2292-2303.
33. Bank, W., *Improving court efficiency: the Republic of Korea's e-court experience*. 2013.
34. Anthes, G., *Estonia: a model for e-government.* Communications of the ACM, 2015. **58**(6): p. 18-20.

35. Danaher, J., *The threat of algocracy: Reality, resistance and accommodation.* Philosophy & Technology, 2016. **29**(3): p. 245-268.
36. Sotala, K. and R.V. Yampolskiy, *Responses to catastrophic AGI risk: a survey.* Physica Scripta, 2015. **90**(1): p. 018001.
37. Yudkowsky, E., *Artificial Intelligence as a Positive and Negative Factor in Global Risk*, in *Global Catastrophic Risks*, N. Bostrom and M.M. Cirkovic, Editors. 2008, Oxford University Press: Oxford, UK. p. 308-345.
38. Armstrong, S., A. Sandberg, and N. Bostrom, *Thinking Inside the Box: Using and Controlling an Oracle AI.* Minds and Machines (to appear), 2012.
39. Yampolskiy, R.V., *Artificial Intelligence Safety and Security*. 2018: CRC Press.
40. Brundage, M., et al., *The malicious use of artificial intelligence: Forecasting, prevention, and mitigation.* arXiv preprint arXiv:1802.07228, 2018.
41. Yampolskiy, R.V. and M. Spellchecker, *artificial intelligence safety and cybersecurity: a timeline of AI failures.* arXiv preprint arXiv:1610.07997, 2016.
42. Babcock, J., J. Kramár, and R. Yampolskiy, *The AGI containment problem*, in *Artificial General Intelligence*. 2016, Springer. p. 53-63.
43. Yampolskiy, R.V. *Taxonomy of Pathways to Dangerous Artificial Intelligence*. in *AAAI Workshop: AI, Ethics, and Society*. 2016.
44. Woodbury, S.A., *Universal Basic Income.* The American Middle Class: An Economic Encyclopedia of Progress and Poverty [2 volumes].(2017), 2017. **314**.
45. Van Parijs, P., *Basic income: a simple and powerful idea for the twenty-first century.* Politics & Society, 2004. **32**(1): p. 7-39.
46. Epstein, R.A., *Citizens United v. FEC: The constitutional right that big corporations should have but do not want.* Harv. JL & Pub. Pol'y, 2011. **34**: p. 639.
47. Bostrom, N., *Existential risk prevention as global priority.* Global Policy, 2013. **4**(1): p. 15-31.
48. Karl, T.L., *Economic inequality and democratic instability.* Journal of Democracy, 2000. **11**(1): p. 149-156.
49. Turing, A., *Computing Machinery and Intelligence.* Mind, 1950. **59(236)**: p. 433-460.
50. Kantardzic, M., et al. *Click Fraud Prevention via multimodal evidence fusion by Dempster-Shafer theory*. in *Multisensor Fusion and Integration for Intelligent Systems (MFI), 2010 IEEE Conference on*. 2010. IEEE.
51. Walgampaya, C., M. Kantardzic, and R. Yampolskiy, *Evidence Fusion for Real Time Click Fraud Detection and Prevention.* Intelligent Automation and Systems Engineering, 2011: p. 1-14.
52. Greenberg, A., *Meet the 'Assassination Market' Creator Who's Crowdfunding Murder with Bitcoins.* Forbes, November, 2013. **18**: p. 2014.
53. Bryson, J.J., M.E. Diamantis, and T.D. Grant, *Of, for, and by the people: the legal lacuna of synthetic persons.* Artificial Intelligence and Law, 2017. **25**(3): p. 273-291.
54. Schweizer, P., *The truly total Turing test.* Minds and Machines, 1998. **8**(2): p. 263-272.
55. Walsh, T., *Turing's red flag.* Communications of the ACM, 2016. **59**(7): p. 34-37.